\documentclass[aps,preprint,superscriptaddress]{revtex4}
\usepackage{revsymb}
\usepackage[dvips]{graphicx}
\usepackage{subfigure}
\usepackage{here}
\begin{document}
\title{Large Global Coupled Maps with Multiple Attractors} 
\author{M. F. Carusela}
\email{flor@ungs.edu.ar} 
\affiliation{Instituto de 
Ciencias, Universidad de General
Sarmiento, Roca 850, 1663 ,San Miguel,
Argentina}
\author{H. Castellini}
\email{hcaste@fceia.unr.edu.ar}
\affiliation{Dpto\@. de F\'{\i}sica, F\@.C\@.E\@.I\@.A\@., 
Pellegini 250,
2000 Rosario}
\author{L. Romanelli}
\email{lili@ungs.edu.ar} 
\affiliation{Instituto de 
Ciencias, Universidad de General
Sarmiento, Roca 850, 1663 ,San Miguel,
Argentina}
\begin{abstract}
A system of N unidimensional global coupled maps (GCM), which support
multiattractors is studied.
We analize the phase diagram and some special features of the transitions
(volumen ratios and characteristic exponents), by controlling the number
of elements of the initial partition that are in each basin of
attraction. It was found important differences with widely known coupled
systems with a single attractor.

\end{abstract}
\maketitle
\section{Introduction}
The emergence of non trivial collective behaviour in multidimensional
systems has been analized in the last years by many
authors \cite{K4} \cite{T1} \cite{Ce1}.
Those important class of systems are the ones that present global
interactions.

A basic model extensively analized by Kaneko is an
unidimensional array of $N$ elements:
\begin{equation}
X_{t+1}(k)=(1-\epsilon)\,f(X_t(k))+\frac{\epsilon}{N}\,
\sum_{l=1}^N f(X_{t}(l))    
\label{eq:sist}
\end{equation}
where $k=1,\ldots,N$,
is an index identifying the elements of the array, $t$ a
temporal
discret variable, $\epsilon$ is the coupling parameter and $f(x)$
describes the local dynamic and taken as the logistic map.
In this work, we consider  $f(x)$ as a cubic map given by:
\begin{equation}
f(x)=(1-a)\,x+a\,x^3 
\end{equation}
where $a \in [0, 4]$ is a control parameter and $x \in [-1, 1]$.
The map dynamic has been extensively studied by Testa et.al.\cite{H1}, 
and many 
applications come up from artificial neural networks where the cubic
map, as local dynamic, is taken into account for modelizing an associative
memory system. Ishi et\@. al\@. \cite{I1} proposed a GCM model 
to modelize this system optimazing the Hopfield's model. 

\begin{figure}[H]
\begin{center}
\mbox{\includegraphics[width=10cm, height=7cm, angle=-90]{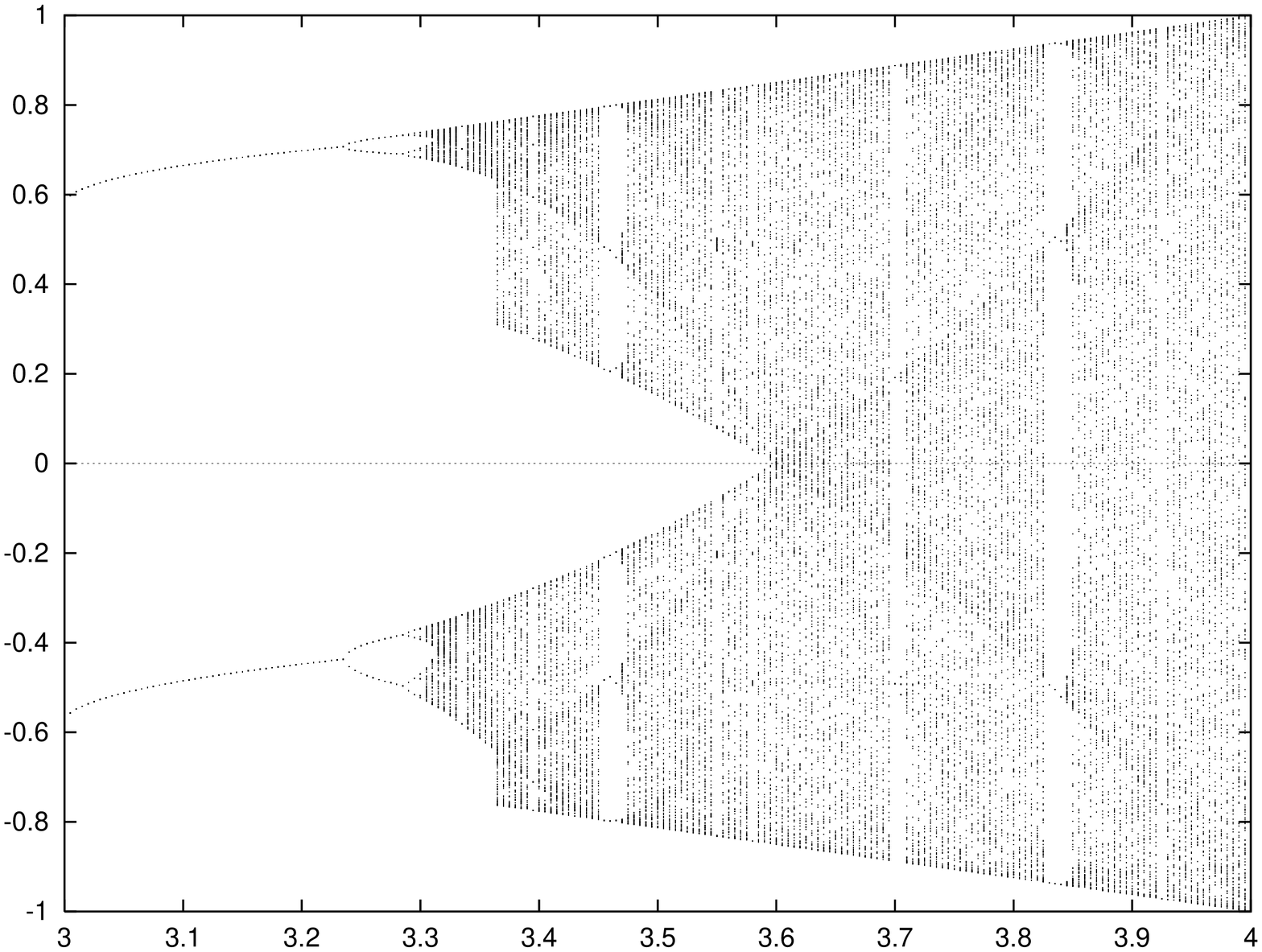}}
\mbox{\includegraphics[width=10cm, height=7cm, angle=-90]{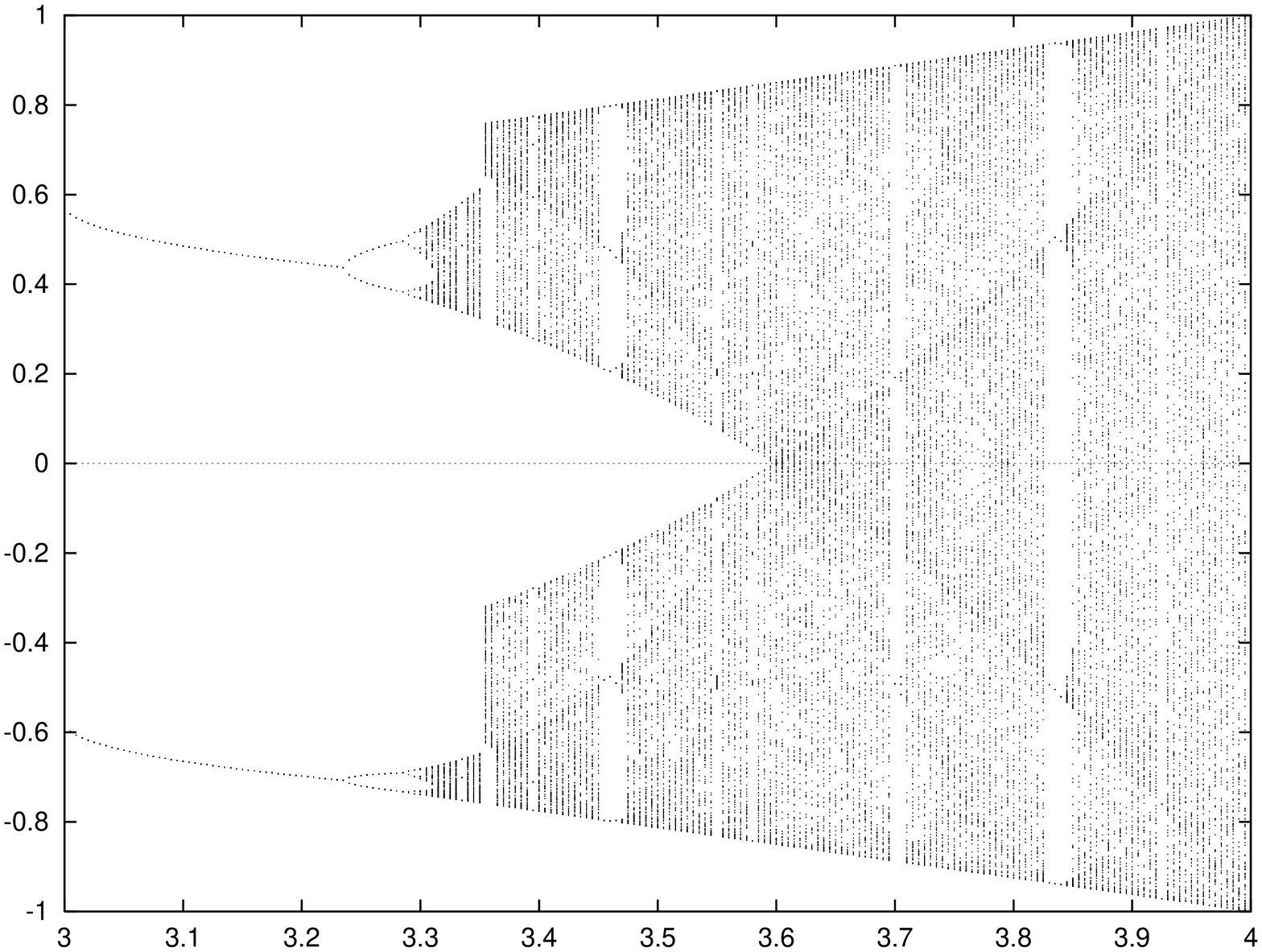}}
\caption{The subharmonic cascade, we graphic 256 iterations of the map versus
$a$ for each basin, after a transient of 5000 steps.}
\end{center}
\label{fig:2}
\end{figure}

The subarmonic cascade, showed on  fig-\ref{fig:2} 
prove the coexistence of
two equal volume stable attractors. The later is verified even as the GCM
given by Eq.\ref{eq:sist} has $\epsilon>0$.
Janosi et\@. al\@. \cite{J1} studied a globally coupled
multiattractor quartic map with different volume basin attractors,  
which is as simple second iterate of the map proposed by Kaneko,
emphazasing their analysis on the control parameter of the local
dynamic. 
They showed that for these systems the mean
field dynamic is controlled by the number of elements in the initial
partition of each basin of attraction. This behaviour is also present in
the map used in this work.

In order to study the coherent-ordered phase transition of the
Kaneko's GCM model, Cerdeira et\@. al\@. \cite{Ce2} analized
the mechanism of the on-off intermitency appearing in the onset of this
transition.
Since the cubic map is characterized by a dynamic with multiple
attractors, the first step to determine the differences with the well
known cuadratic map given by Kaneko is to obtain the phase diagram of
Eq.\ref{eq:sist} and to study the the coherent-ordered dynamical
transition for a fixed value of the control parameter $a$. 
The later is done near an internal crisis of the cubic map,
as a function of the number of elements $N_1$
with initial conditions in one basin and the values of
the coupling parameter $\epsilon$, setting $N$ equal to 256.         
After that, the existence of an inverse period
doubling bifurcation as function of $\epsilon$ and $N_1$ is analized.  

\section{Phase Diagram}
\begin{figure}[H]
\centering
\mbox{\includegraphics[width=8cm, height=8cm]{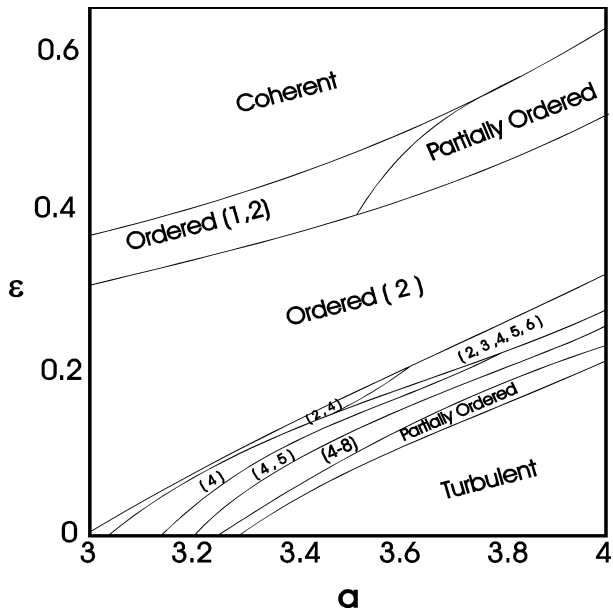}}
\caption{Phase diagram as function of $\epsilon$ and $a$. Number such 
$(2,4)$ represents dominant cluster
number (that is with basin volume ratio more than 10\%)}
\label{fig:1}
\end{figure}

The dynamical analysis process breaks the phase space in sets  
formed by synchronized elements which are called clusters. 
This is so, even when, 
there are identical interactions between identical elements.
The system is labeled as {\em 1-cluster}, {\em 2-cluster}, etc\@. state
if
the $X_i$ values fall into one, two or more sets of synchronized elements
of the phase space.
Two different elements $i$ and $j$ 
belong to the same cluster within a
precision $\delta$ (we consider $\delta=10^{-6}$) only if  
\begin{equation}
\left|X_i(t)-X_j(t)\right|\leq\delta  
\end{equation} 
Thus the system of Eq.\ref{eq:sist}, shows the existence of
different phases with clustering 
(coherent, ordered, partially ordered, turbulent).
This phenomena appearing in GCM was studied by Kaneko 
for logistic coupled maps when the
control and coupling parameters vary. 
A rough phase diagram for an array of 256 elements is determined for the
number of clusters calculated from 500 randomly sets of initial conditions
within the precision specified above.
This diagram displayed in fig-\ref{fig:1}, was
obtained following the criteria established by this author.
Therefore, the  $K$ number of clusters and the number of elements that
build them are relevant magnitudes to characterize the system
behaviour.

\section{Phase Transition}
\begin{figure}[H]
\centering
\mbox{\includegraphics[width=8cm, height=8cm, angle=-90]{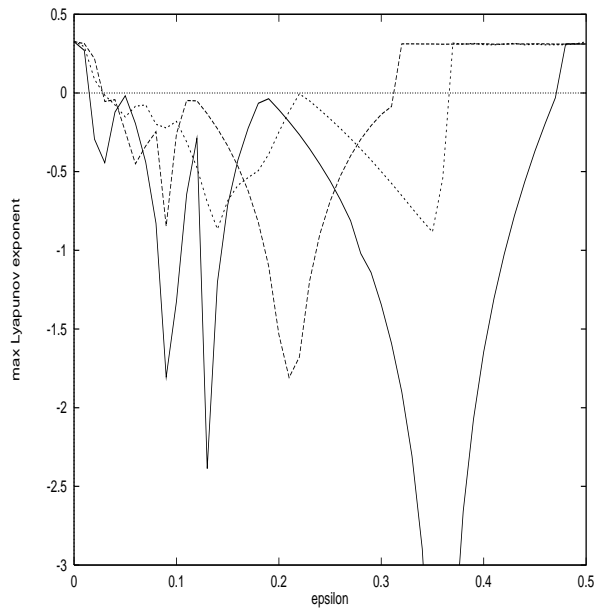}}
\caption{Maximum Lyapunov exponents for three different values of $N_1$.
The solid line corresponds to $N_1=128$, 
the dashed line to $N_1=95$, and the doted line to $N_1=64$.}
\label{fig:4}
\end{figure}

\begin{figure}[H]
\centering
\mbox{\includegraphics[width=8cm, height=8cm, angle=-90]{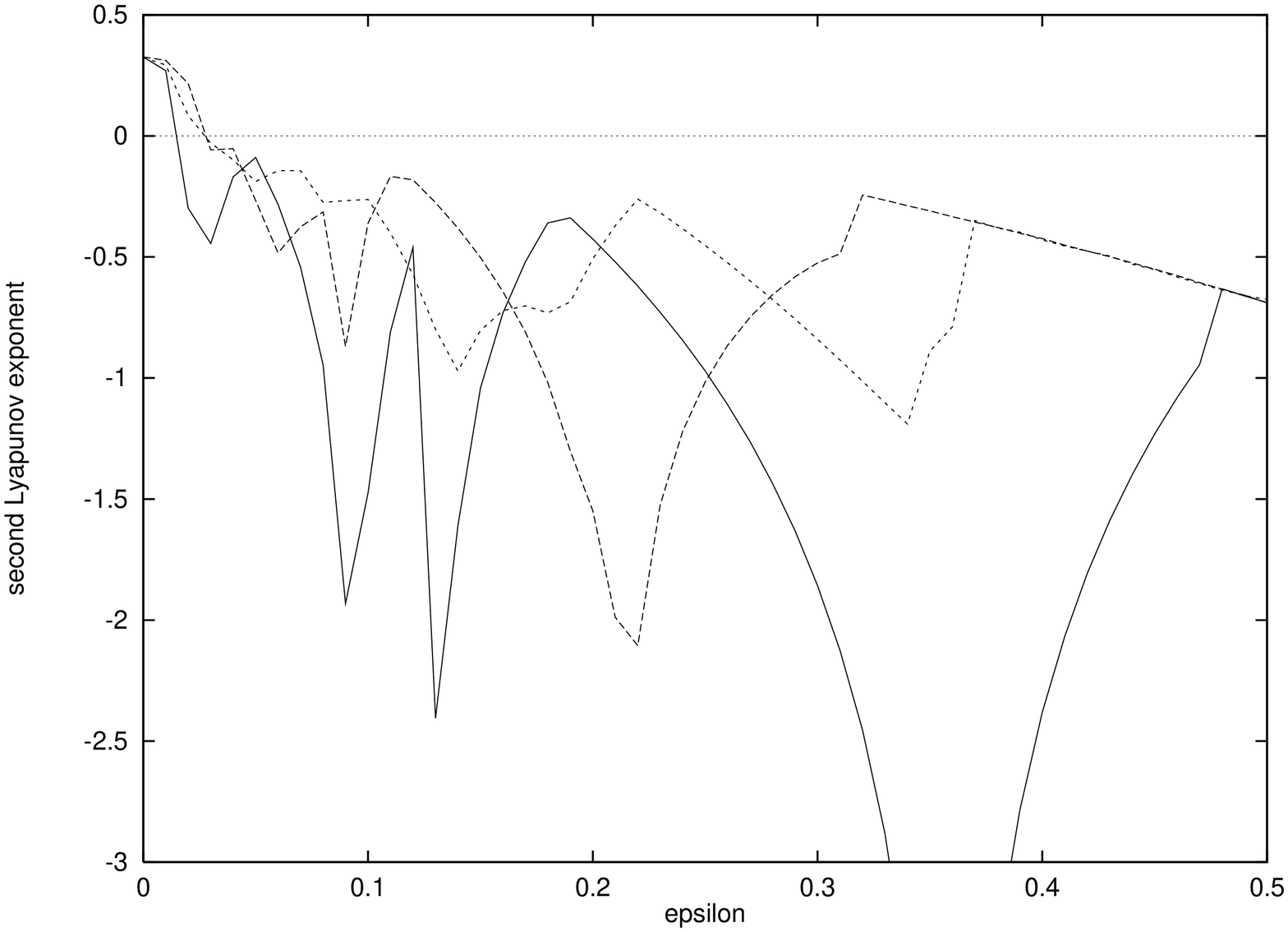}}
\caption{Second Lyapunov exponents for three different values of $N_1$ 
The solid line corresponds to $N_1=128$, 
the dashed line to $N_1=95$, and the doted line to $N_1=64$.}
\label{fig:5}
\end{figure}

In order to study phase transition, the two greatest Lyapunov exponents
are shown in fig-\ref{fig:4} and fig-\ref{fig:5}. 
They are depicted for a=3.34 as a function of $\epsilon$ and for three
different values 
of initial elements $N_1$.

\begin{figure}[H]
\centering
\mbox{\includegraphics[width=8cm, height=8cm]{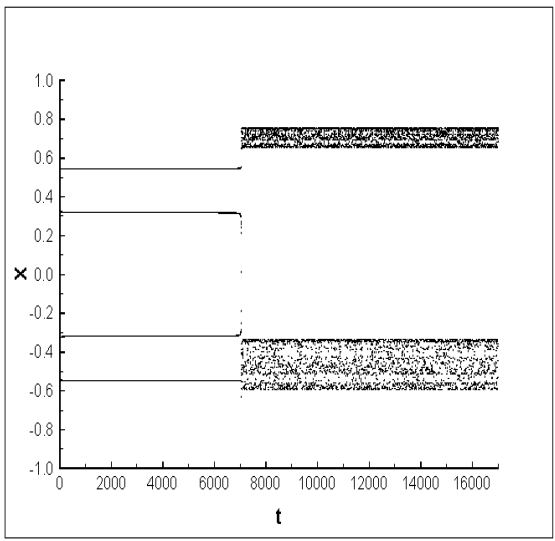}}
\caption{$X$ vs\@. $t$ for two elements of differents clusters near the 
metastable transition, for $a=3.34$, $N_1=95$ and $\epsilon=0.3697126$.}
\label{fig:3}
\end{figure}

\begin{figure}[H]
\centering
\mbox{\includegraphics[width=8cm, height=8cm, angle=-90]{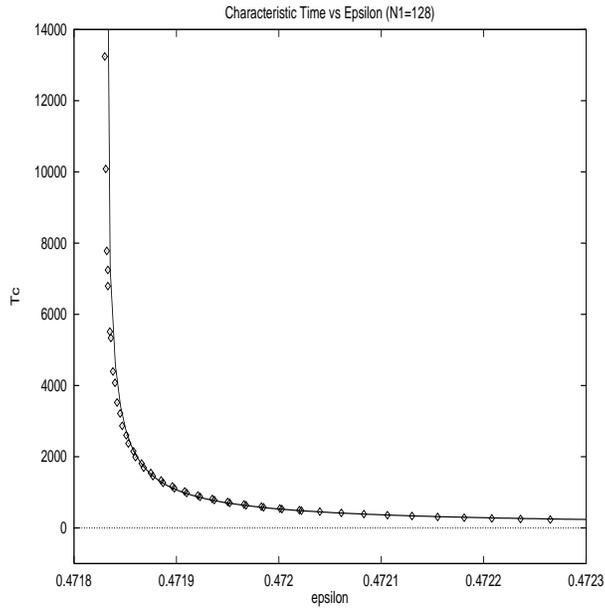}}
\caption{Characteristic time $T_t$ for $N_1=128$, doted line, and its
fitting with the Eq.\ref{eq:1}, solid line.}
\label{fig:6}
\end{figure}

In the coherent phase, as soon as $\epsilon$ decrease,
the maximum Lyapunov exponent changes steeply from a positive to a
negative 
value when the two cluster state is reached. A 
sudden change in the attractor phase space occurs for a critical value
of the coupling parameter $\epsilon_c$  in the analysis of the transition
from two to one cluster state. Besides that, in the same transition for
the same $\epsilon_c$, a metastable transient state of two cluster to one
cluster chaotic state is observed, due to the existence of an unstable
orbit inside of the chaotic basin of attraction, as is shown in
fig-\ref{fig:3}
The characteristic time $T_t$ in which the system is entertained in
the
metastable transient is depicted in Fig-\ref{fig:6}, for values of 
$\epsilon$ near and above $\epsilon_c$.

For a given set of initial conditions, it is possible to fit this
transient as:
\begin{equation}
T_t=(\frac{\epsilon-\epsilon_c}{\epsilon_c})^{-\beta}
\label{eq:1}
\end{equation}
This fitting exponent $\beta$, depends upon the number of elements with 
initial
conditions in each basin as is shown in the next table for three
$N_1$ values and setting $N=256$.

\begin{center}
\begin{tabular}{|l|l|l|}
\hline
$N_1$ & $\epsilon_c$ & $\beta$ \\
\hline
128 & 0.471829 & 0.792734 \\
95 & 0.3697115 & 0.606751 \\
64 & 0.3198161 & 0.519833 \\
\hline
\end{tabular}
\end{center}

It is worth noting from the table that $\beta$ increases with 
$N_1$ up to $N/2$, and for $N_1$ due to the basins symmetry. 

\section{Inverse Period Doubling Bifurcation}
\begin{figure}[H]
\begin{center}
\includegraphics[width=8cm, height=8cm]{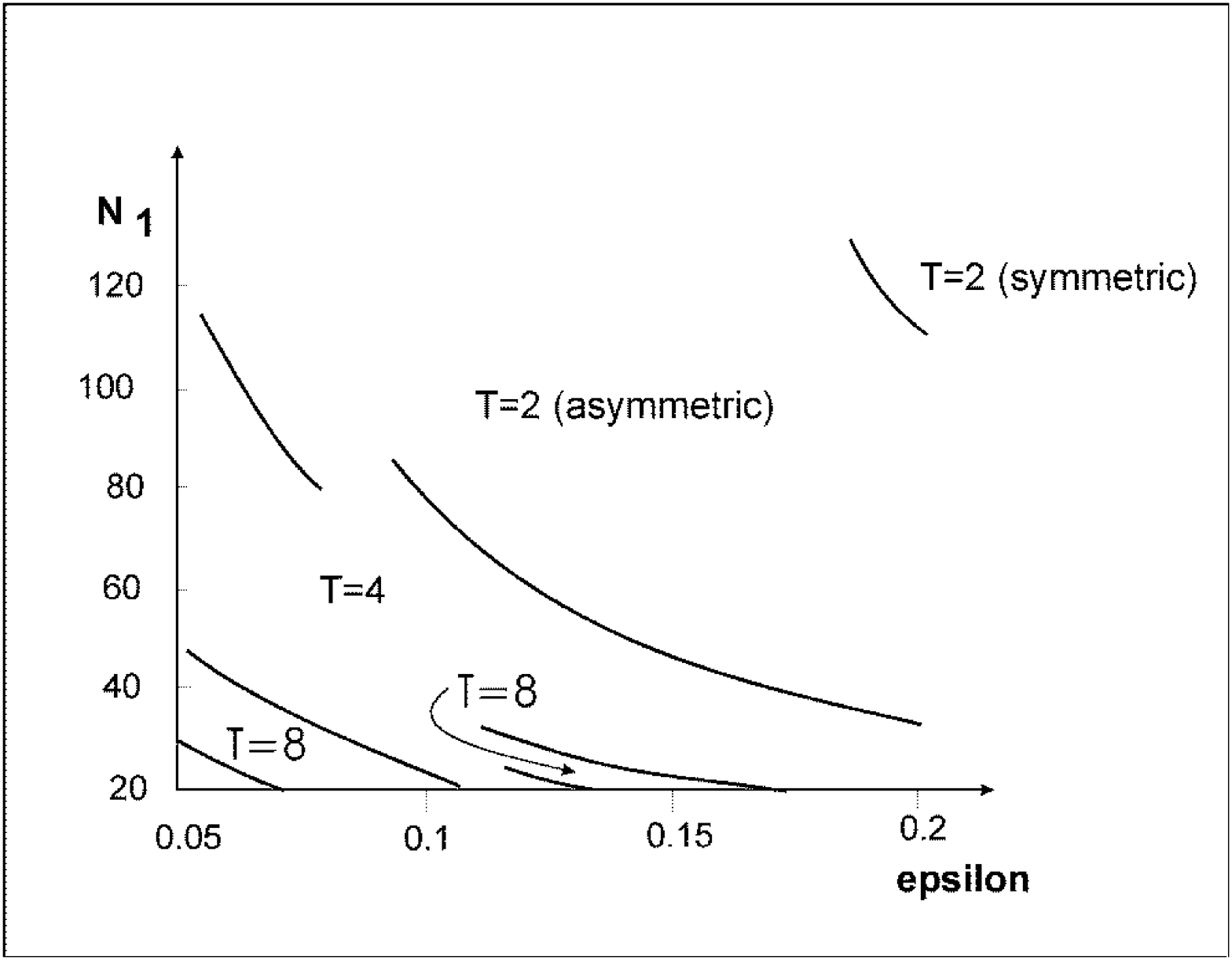}
\caption{Bifurcation diagram for $a=3.34$ up to period 8, obtained from 1000
sets of initial conditions.}
\label{fig:7}
\end{center}
\end{figure}

In order to analize the existence of period doubling bifurcations, the
maxima Lyapunov exponent $\lambda_{max}$ is calculated as function of $N_1$
and $\epsilon$. For each $N_1$, critical values of the coupling
parameter, called $\epsilon_{bif}$, are observed when a negative
$\lambda_{max}$ reaches a zero value without changing sign. This behaviour is
related to inverse period doubling bifurcations of the GCM.
Fitting all these critical pair of values $(\epsilon_{bif}, N_1)$, a
rough $N_1$ vs $\epsilon_{bif}$ graph is shown in fig-\ref{fig:7}, 
and different
curves appears as boundary regions of the parameter space where
the system displays $2^n$ ($n=1, 2, 3$) periods states .
This is obtained without taking into accout the number of final clusters. 
It is clear that greater values of $N_1$, correspond to smaller
$\epsilon_{bif}$
for the occurrence of the bifurcation.
Evidence of period 16 appears for values of $N_1 $ smaller than 30.
In fig-\ref{fig:7} T=2(symmetric) means period two orbit, with
clusters oscillating with equal amplitud around zero, 
T=2(asymmetric) means period two orbit, with clusters
oscillating with different amplitud.  

\section{Concluding Remarks}
The study of systems with coexistence 
of multiple attractors gives a much richer dynamics and a new control
parameter must necessarily be added. 
Although the dimensionality in the parameter space is increased by one, 
the dynamics is
rather simple to characterize. 
Some of the relevant aspects of this kind of systems are shown in this
work.
The phase diagram that was obtained shows the existence of similar phases 
to those using the cuadratic and quartic map, 
this behaviour suggests some kind of
universality in the dynamics of the GCM.
Another interesting issue found, concerns the metastable transition
between two to one cluster state, along with 
a sudden jump in the maximum Lyapunov exponent, as it was displayed in
fig.\ref{fig:7}.
The characteristic time given by Eq.\ref{eq:1} also correspond to the
above
transition where the critical exponent $\beta $ and the 
critical coupling parameter $\epsilon $ shows a strong dependence 
on the number of initial elements in each basin.
An inverse bifurcation cascade appears 
when the system is in two or more clusters state where $\epsilon $ 
and $N_1 $ are the 
critical parameters of the bifurcation, 
which means the maximum Lyapunov exponent is equal to zero. 

\section{Acknowledgments}
 This work is partially supported by CONICET
(grant PIP 4210). MFC and LR also express their acknowledgment to the ICTP
where the 
initial discussion of the work was performed.








\end{document}